# High-order (*N*=4–6) multiphoton absorption and mid-infrared Kerr nonlinearity in GaP, ZnSe, GaSe, and ZGP crystals.

Taiki Kawamori,[1] Peter G. Schunemann,[2] Vitaly Gruzdev,[3] and Konstantin L. Vodopyanov[1,a]

1. CREOL, College of Optics and Photonics, University of Central Florida, Florida 32816, USA
2. BAE Systems, P. O. Box 868, MER15-1813, Nashua, New Hampshire 03061-0868, USA
3. Department of Physics and Astronomy, University of New Mexico, Albuquerque, New Mexico 87106 USA

a) *vodopyanov@creol.ucf.edu*

## ABSTRACT

We report a study of high-order multiphoton absorption, nonlinear refraction, and their anisotropy in four notable mid-infrared $\chi^{(2)}$ crystals: GaP, ZnSe, GaSe and ZGP using the Z- scan method and 2.35-µm femtosecond pulses with peak intensity in excess of 200 GW/cm$^2$. We found that the nonlinear absorption obeys a perturbation model with multiphoton absorption (MPA) orders from $N$ = 4 to 6, in good agreement with the bandgaps of the crystals. A study of the role of free carrier absorption, performed by varying the pulse duration between 30 and 70 fs while maintaining a constant peak intensity showed that at our intensity levels, free carriers generated in the process of MPA absorb much stronger than would be expected from their linear absorption cross section. Possible mechanisms include high-field effects, such as intravalley scattering in the conduction band and absorption to higher lying bands. Nonlinear refractive indices were measured using (i) closed aperture Z-scan and (ii) spectral broadening due to self-phase modulation, both methods agreeing well with each other.

## I. INTRODUCTION

With the ability to generate few-cycle optical pulses from mode-locked lasers, focused power densities of >100 GW/cm$^2$ are now achievable even at full (~100 MHz) repetition rates using oscillator outputs. The development of ultrafast sources in the mid-infrared range (MIR, $\lambda$ > 2 µm) made it possible to demonstrate new high-field effects, such as generation of high harmonics in solids [1,2,3,4], producing multi-octave supercontinuum [5,6,7,8], producing ultra-broadband frequency combs via subharmonic generation [9,10], and generating single-cycle MIR transient via optical rectification [11,12]. The main motivation of his work was to study the behavior of MIR non-linear materials in terms of high-order detrimental effects such as multiphoton absorption.

The crystals studied here, GaP, ZnSe, GaSe, and ZGP (ZnGeP$_2$), are notable MIR materials with high quadratic nonlinearity and broad MIR transparency extending beyond 10 µm. Thanks to their relatively large bandgap energy ($\geq$ 2 eV, see Table 1), they do not suffer from two- or three-photon absorption when pumped at $\lambda$>1.9 µm. However, with increasing irradiance through the use of e.g. few-cycle pulses, high-order effects come into play. At the same time, literature data on high-order ($N$> 3) multiphoton absorption (MPA) are uncommon, with the exception of [13] reporting four-photon absorption in GaP. Equally important is the knowledge of MIR nonlinear refractive indices that is critical for supercontinuum and Kerr frequency comb generation as well as for self-focusing effects.

Here we report our experimental results on MPA and nonlinear refraction measurements in GaP, ZnSe (single and polycrystalline), GaSe, and ZGP using ultrashort pulses. We measure MPA anisotropy, derive MPA coefficients, compare them with theory, and study the role of free carriers produced during the MPA process.

## II. EXPERIMENTAL SETUP

A schematic diagram of the experiment is shown in Fig.1. We characterized the crystals using the open- and closed-aperture Z-scan [14]. A beam of a linearly polarized Kerr-lens mode-locked 2.35 µm Cr$^{2+}$:ZnS laser with pulse duration 62 fs, pulse repetition rate 79 MHz, and the average power 1.2 W was focused by a 90-degree off-axis parabolic mirror (OAP), and the transmitted power was recorded while the sample was translated along the focus by a motorized stage. The samples were slightly tilted with respect to the normal incidence to avoid back reflection (with exception of ZGP, the samples were not anti-reflection coated). The transmitted signal was measured with a large-area pyroelectric detector. The input intensity was controlled by a variable neutral density filter (2-mm fused silica substrate) and monitored by a reference detector. The beam waist ($1/e^2$ intensity radius) was measured with the knife-edge technique and was found to be $w$=8.5 μm for $f$ = 15 mm OAP and $w$=12 μm for $f$ = 25 mm OAP, with the beam quality factor $M^2$ <1.1. Figs. 1(b-d) show the second-order autocorrelation trace for the laser pulses, the laser spectrum, and measured beam size near the focus for the $f$=15 mm focusing mirror.

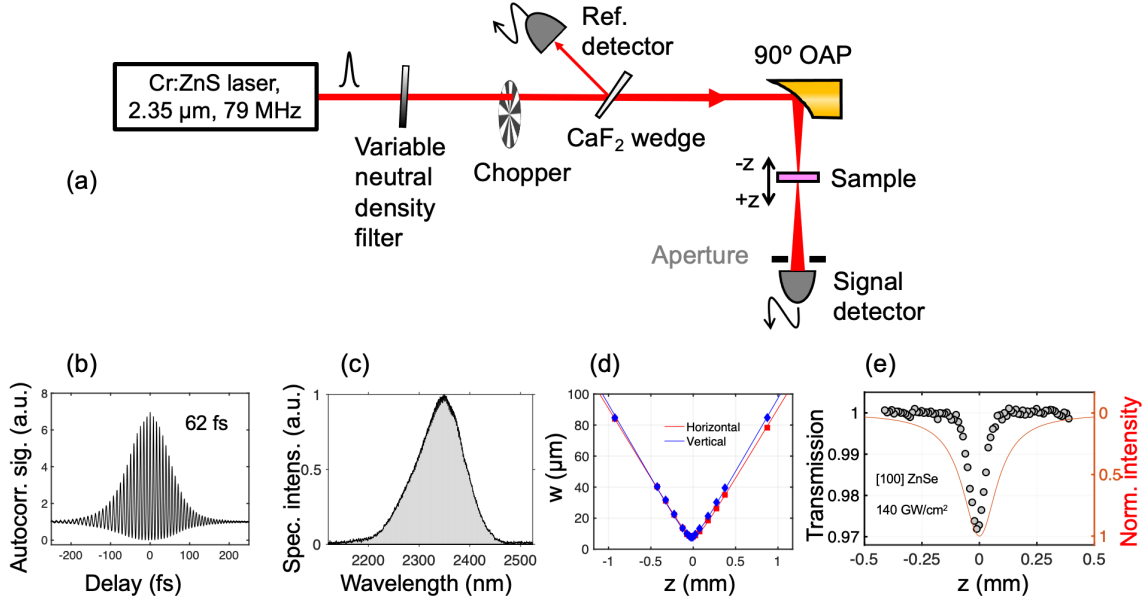

FIG. 1. (a) Experiment setup for Z-scan measurements. (b) Second-order autocorrelation trace of a laser pulse. (c) Laser spectrum. (d) Measured beam size near focus for $f$=15 mm OAP. (e) Nonlinear transmission of a 0.1-mm ZnSe crystal in the open aperture Z-scan (circles) along with peak-on-axis intensity variation along the beam (solid line, inverted).

Table 1 lists some properties of the crystals under study, which include the theoretical MPA order $N$, defined as the nearest integer above $E_g/\hbar\omega$ ratio, where $\hbar\omega$ is the photon energy at λ = 2.35 μm and $E_g$ is the bandgap.

Table 1. Properties of the four crystals under study.

| Crystal | Transparency range (μm) | 2nd order nonlinearity (pm/V) | Bandgap energy (eV) | MPA order $N$ at λ=2.35μm (theory) |
|---|---|---|---|---|
| GaP | 0.57–13 | $d_{14}$=35 | 2.79 (direct) | 6 |
| ZnSe | 0.55–20 | $d_{14}$=20 | 2.7 | 5-6* |
| GaSe | 0.65–20 | $d_{22}$=54 | 2.0 | 4 |
| ZGP | 0.74–12 | $d_{36}$=75 | 2.0 | 4 |

*) The laser spectrum falls between 5- and 6-photon absorption for ZnSe.

## III. MULTIPHOTON ABSORPTION

### A. Theoretical considerations

For MPA of the order $N$, the beam intensity evolution along the propagation direction $z$ is described by by the following coupled nonlinear equations:

$$\frac{dI}{dz} = -\alpha_N I^N - \sigma_c n_c I, \qquad (1)$$

$$\frac{dn_c}{dt} = \frac{\alpha_N I^N}{N\hbar\omega} - \frac{n_c}{\tau_c}. \qquad (2)$$

Here $I = I(r, t, z)$ is the local intensity of the beam, $\alpha_N$ is the $N^{th}$ order MPA coefficient, $\hbar\omega$ is the photon energy, $n_c$ is the density of free carriers accumulated during the MPA process, $\sigma_c$ is the free carrier absorption (FCA) cross section, and $\tau_c$ is the carrier lifetime. We assume that only one type of MPA dominates for a given laser spectrum. From (1-2) it follows that

$$\frac{dI}{dz} = -\alpha_N I^N - \sigma_c I \int_0^t \frac{\alpha_N I(t')^N}{N\hbar\omega} dt' \qquad (3)$$

The first term on the right side of this equation (due to MPA) is proportional to $I^N$, while the second term (due to FCA) – to $I^{N+1}$. Consequently, the former will prevail at low peak intensities, when the normalized transmittance change $\Delta T = \Delta I/I$ tends to zero; it also follows from (3) that the shorter the pulse duration, the less the role of FCA. We assume that the free carrier lifetime is on the order of a few ns [15] and is shorter than the pulse repetition period, so there are no carriers left from the previous pulse. Assuming $|\Delta T| \ll 1$ ($I \approx const$) and neglecting FCA, one can write

$$\frac{dI}{dz} = -\alpha_N I^N , \qquad (4)$$

and, integrating over the $z$ coordinate, we obtain

$$\Delta T = \Delta I/I = -\alpha_N L I^{N-1} , \qquad (5)$$

where $L$ is the sample length. The slope of the *log-log* plot of $|\Delta T|$ vs. peak intensity is expected to be $N$-1, which can be used to determine the multiphoton order $N$.

To address the influence of diffraction and self-focusing inside the sample for quantitative calculations of both nonlinear refraction and MPA coefficients, we used the nonlinear Schrödinger equation in the form of [16] for the complex field amplitude, modified to take into account high-order MPA and FCA:

$$\frac{\partial A}{\partial z} = \frac{i}{2k} \Delta_\perp A - \frac{1}{2}\alpha_N |A^2|^{(N-1)} A + i\frac{\omega}{c} n_2 |A^2| A - \frac{1}{2}\sigma_c n_c A. \qquad (6)$$

Here $A = A(r, t, z)$ is the normalized complex electric field amplitude (such that $|A^2| = I$), $k$ is the wavenumber, $\Delta_\perp$ is the transverse Laplacian operator, and $n_2$ is the nonlinear refractive index. The overall transmittance of the sample was calculated by temporal and spatial (across the beam) integration of the output electric field intensity.

**B. MPA experimental results**

To study MPA and determine its predominant order $N$, we performed the open-aperture Z-scan with a fixed pulse duration ($\tau$=62 fs) and OAP mirror with $f$ = 15 mm. Fig. 1(e) shows a raw Z-scan result with a 100-μm-thick monocrystalline ZnSe sample oriented along [100] with respect to the laser beam polarization.

Fig. 2 shows logarithmic plots of the Z-scan dip $\Delta T$ versus peak peak-on-axis intensity at the focus inside the sample. The incoming laser power was varied with a neutral density filter. For all four crystals under study we observed the linear *log-log* scale dependence with a constant slope. For GaP the slope varied from 4.6 to 6.5 depending on the crystal orientation with the average of 5.6, at the expected slope of $N$-1=5 (Table 2) for the direct bandgap of GaP. We also compared bulk GaP and its orientation-patterned version (OP-GaP) – a structure with periodically inverted crystalline domains used for quasi-phase-matched applications in nonlinear optics [17]. A 700-μm layer of OP-GaP was grown by the hydride vapor phase epitaxy on a 350-μm-thick (100) GaP substrate. We toggled the beam between patterned and un-patterned portions of GaP and did not observe any difference in MPA. For ZnSe, the slope was close to 4 that indicates that the $N$=5 MPA process was dominant. In fact, the laser spectrum falls between 5- and 6-photon absorption for ZnSe, and about 20% of the area of the laser spectrum is at λ<2.3 μm, corresponding to the 5-photon absorption [Fig.1(b)]. It appears that the contribution of the $6^{th}$ order MPA was small at our intensities. For the polycrystalline ZnSe, no power law dependence was observed (the sample used was the same as in [18] with the average grain size of 100 μm). Rather, the *log-log* slope varied from 0.6 to 2 when the intensity varied from 10 to 100 GW/cm$^2$ [inset to Fig. 2(b)]. This behavior may be explained by possible existence of intermediate energy states in the forbidden gap of polycrystalline, ZnSe and due to admixture of the $2^{nd}$, $3^{rd}$ and $4^{th}$ optical harmonics produced via random phase matching [19]. The GaSe and ZGP crystals showed power law dependence with the *log-log* slope close to 3 corresponding to the expected 4-photon absorption.

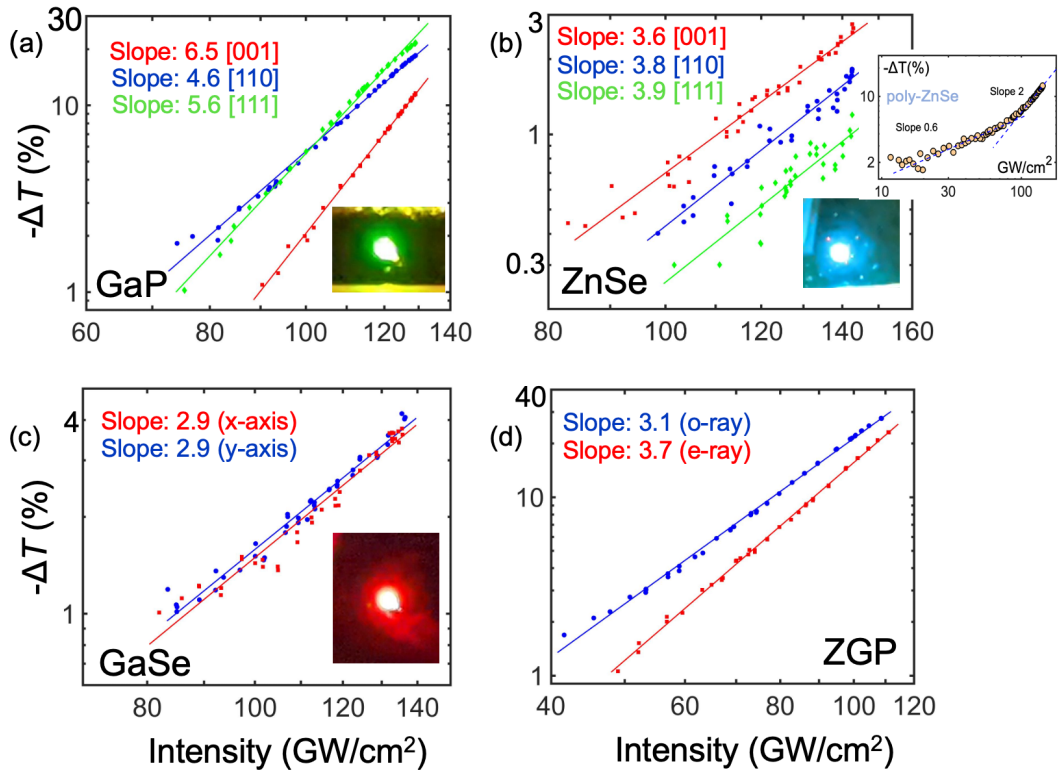

FIG. 2. Logarithmic plots of the Z-scan transmission dip $\Delta T$ vs. peak-on-axis intensity at the focus ($\tau$=62 fs pulses) for the four crystals studied. The insets in (a-c) show the interband photoluminescence from the crystals.

Experimental values for the MPA orders $N$ based on the linear slope fitting – for different crystals and their orientations, are summarized in Table 2. There is a good match (to within -0.2…+0.6) between experimental MPA orders (averaged over crystals' orientations) with the ones derived from the direct bandgaps of the crystals.

Table 2. MPA orders at $\lambda$=2.35 μm, theory vs. experiment. MPA coefficients and peak intensity corresponding to the onset of MPA for each crystal.

| Crystal | Sample length (mm) | Crystalline direction | MPA order $N$ (theo.) | MPA order $N$ (exp.) | Ave. $N$ (exp.) | Peak intensity ($\tau$=62 fs) for MPA to reach 1 cm$^{-1}$ (GW/cm$^2$) |
|---|---|---|---|---|---|---|
| GaP, 110-cut | 0.5 | [001] [110] [111] | 6 | 7.5 5.6 6.6 | 6.6 | 115 100 100 |
| ZnSe, 110-cut | 0.1 | [001] [110] [111] | 5-6 | 4.6 4.8 4.9 | 4.8 | 110 125 140 |
| GaSe, 001-cut | 0.46 | [100] [010] | 4 | 3.9 3.9 | 3.9 | 150 150 |
| ZGP, θ=50°,φ=0 (approx. 101-cut) | 1 | [010]* [101]** | 4 | 4.1 4.7 | 4.4 | 80 90 |

* o-ray; ** e-ray

The last column of Table 2 shows the MPA thresholds for 62 fs pulses, defined as the peak on-axis intensity when the integrated (in time and in the transverse plane) absorption coefficient reaches 1 cm$^{-1}$. Although we are comparing crystals with different MPA orders, this comparison can be a useful guide for practical applications.

The insets in fig. 2(a-c) show the interband photoluminescence from the crystals caused by the recombination of generated carriers. ZnSe had blue and GaSe – red light emission. Although GaP has a larger band gap than ZnSe, it emits at a longer (green) wavelength due to recombination through its indirect bandgap of 2.24 eV.

## C. The role of free carriers

In contrast to the MPA, which depends only on the peak intensity, the FCA increases with increasing pulse duration due to the accumulation effect. We studied the role of FCA by using another mode-locked 2.35-µm laser with shorter, 30 fs, pulse duration. We varied the pulse duration from 30 to 70 fs by frequency chirping the pulses with dispersive sapphire plates. The result of chirping on duration is shown in Fig. 3(a).

Fig. 3(b) shows the Z-scan dip |$\Delta T$| versus pulse duration dependence at a fixed intensity. The nonlinear absorption strongly increases with pulse duration. A simple model as described in Eq. (3) based on the known FCA cross sections that assumes that one electron-hole pair is created for every $N$ absorbed photons and that FC absorption is dominated by electrons in the conduction band – does not account for this increase. Dashed and dotted curves in Fig. 3(b) represent simulated dependencies for GaP and ZnSe respectively, using equations (1-2) and literature data for FCA cross-sections ($\sigma_c$) at $\lambda \approx 2.35$ µm: $10^{-16}$ cm$^2$ for GaP [20] $2.7 \times 10^{-18}$ cm$^2$ for ZnSe [21].

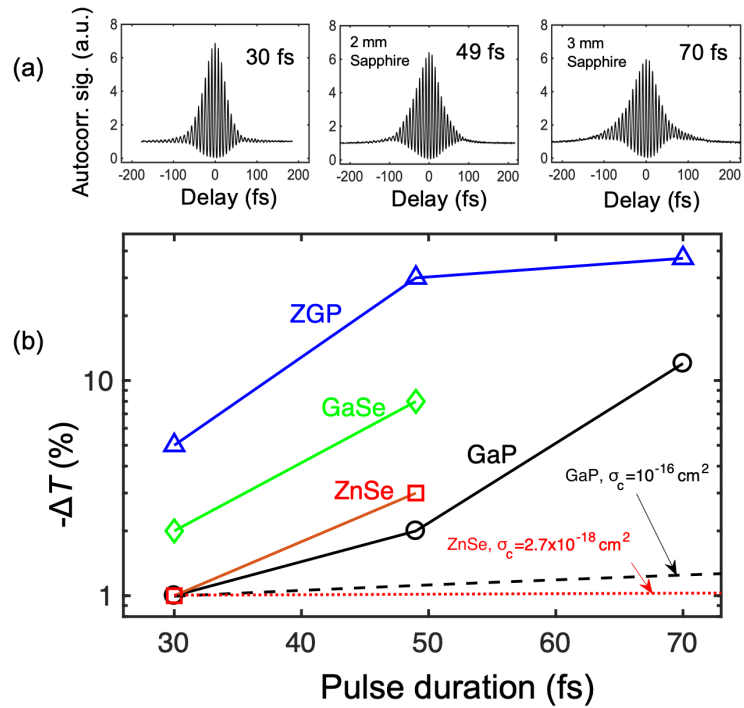

FIG. 3. (a) 2$^{nd}$-order autocorrelation traces for the pulses with progressively increasing duration achieved via frequency chirping of the 30-fs laser pulses. (b) The Z-scan dip $\Delta T$ versus pulse duration at a fixed on-axis peak intensity $I_0$. The intensities and crystal orientations were: 103 GW/cm$^2$ ([111] GaP), 217 GW/cm$^2$ ([111] ZnSe), 115 GW/cm$^2$ ([010] ZGP), and 230 GW/cm$^2$ ([010] GaSe). Dashed and dotted curves represent simulated dependencies for GaP and ZnSe respectively, based on their known free carrier linear absorption cross-sections.

We hypothesize that, at our peak intensities, high field effects may be responsible for the much stronger absorption by free carriers. Consider the example GaP, where the carriers are promoted to the conduction band via 6$^{th}$ order MPA process, about 0.26 eV above the conduction band minimum at zone center, which creates non-thermalized distribution of carriers (Fig.4). These carriers, with the additional field-induced acceleration (the peak ponderomotive energy corresponding to 100 GW/cm$^2$ is 0.23 eV) can scatter to the neighboring conduction-band valley, separated from the band minimum by a 0.51-eV barrier (Fig. 4). Yet another possibility in GaP is direct excited state absorption to the higher conduction band as shown in Fig. 4.

We hypothesize that at our peak intensities, high-field effects may be responsible for the much stronger absorption by the free carriers. For example, in the case of GaP, the six-photon excitation from the heavy-hole valence band creates a non-thermalized distribution 0.28 eV above the bottom of the Γ conduction-band valley (Fig. 4). In addition, the acceleration of the free electrons by the laser-pulse electric field supplies the peak ponderomotive energy of about 0.26 eV at the peak intensity of 100 GW/cm$^2$. The total energy of those free electrons (about 0.54 eV) exceeds the energy barrier of 0.51 eV that separates the Γ and X valleys of the lowest conduction band and creates conditions for scattering to the X valley. Also, the accelerated free electrons can make single-photon transitions to the upper conduction band as shown in Fig. 4.

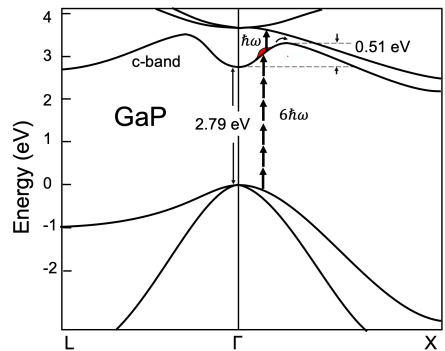

FIG. 4. The band diagram of GaP. Shown is a 6-photon transition from the valence band to the conduction band, as well as direct transitions to the 2$^{nd}$ conduction band due to nonthermalized distribution of carriers.

## IV. NONLINEAR REFRACTION

For the nonlinear refractive index ($n_2$) measurements, we have used 62-fs laser pulses and $f$=25 mm OAP in the closed-aperture Z-scan configuration (Fig. 1). Fig. 5 shows Z-scan traces with the numerical fitting from Eq. (6) without MPA and FCA terms. The latter assumption is justified because the laser intensity was kept low, so that the effects of MPA and FCA were negligible. The results for the nonlinear refraction index data are shown in Table 2.

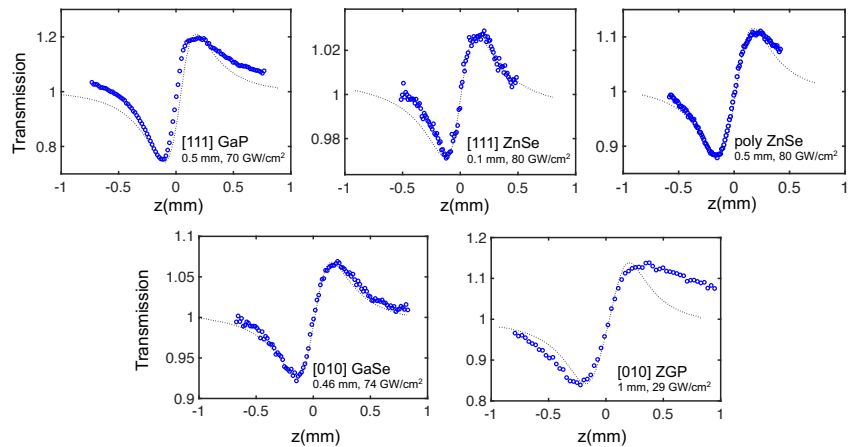

FIG. 5. Experiment close-aperture Z-scan traces taken with 62-fs pulses. The dotted lines are theoretical based on Eq. (6) and $n_2$ as a fitting parameter. The insets show the crystal orientation, length, and on-axis peak intensity.

We complemented our nonlinear refractive index measurements with spectral broadening experiments performed at higher peak intensities. With the samples placed at the beam focus, the output spectrum was measured with a laser spectrum analyzer (Bristol Instruments model 771B). The results for the self-phase modulation (SPM) induced spectral broadening in four crystals are shown in Fig. 6. The peak intensities within the samples varied from 138 GW/cm$^2$ (GaP) to 185 GW/cm$^2$ (ZGP). For the monocrystalline ZnSe, no spectral broadening was observed because of the sample's small thickness. To extract n2 from these measurements, we modeled the spectral broadening using Eq. (6), with the presence of MPA and without the FCA. The spectrum of the pulse after the nonlinear propagation was analyzed in frequency domain by Fourier transform of the complex field electric field. For GaP, the SPM broadened spectrum was asymmetric and did not exactly fit to the model; we estimated $n_2$ by comparing the spectral widths (at half maximum) for the incoming and outcoming pulses, with an estimated uncertainty of ±40%. Within this margin of error, the Z-scan and the spectral broadening data are in good agreement with each other (Table 2). There is also a good agreement with the literature data for GaP and ZnSe, but a greater discrepancy for GaSe and ZGP (our $n_2$ values are about 3 times larger).

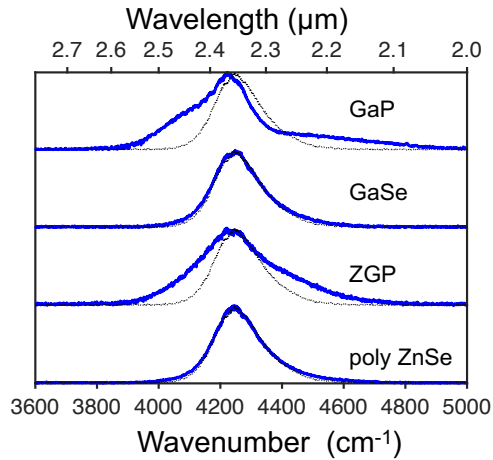

FIG. 6. Results for the SPM-induced spectral broadening in four crystals. The peak irradiance within the samples and their orientation with respect to the laser polarization were: 138 GW/cm$^2$ ([111] GaP); 145 GW/cm$^2$ ([010] GaSe); 185 GW/cm$^2$ ([010] ZGP); and 152 GW/cm$^2$ (poly ZnSe).

Table 2. Nonlinear refraction index data.

| Crystal | n2 from literature (m$^2$/W) | Ref. | n2 (2.35µm) from Z-scan (m$^2$/W) | n2 (2.35µm) from spectral broadening (m$^2$/W) |
|---|---|---|---|---|
| GaP | 3.5x10$^{-18}$ (1.55 µm)<br>1.9 x 10$^{-18}$ (1.75 µm) | [22]<br>[13] | (1.8±0.6)x10$^{-18}$ [100]<br>(1.8±0.6)x10$^{-18}$ [110]<br>(1.8±0.6)x10$^{-18}$ [111] | (1.8±0.7) x 10$^{-18}$ [111] |
| ZnSe | 1.5x10$^{-18}$ (2.0 µm)<br>1.1x10$^{-18}$ (2.3 µm)<br>1.2x10$^{-18}$ (3.9 µm) [poly]<br>1.2x10$^{-18}$ (10 µm) [poly] | [23]<br>[24]<br>[25]<br>[26] | (1.2±0.4)x10$^{-18}$ [100]<br>(1.0±0.4)x10$^{-18}$ [110]<br>(1.0±0.4)x10$^{-18}$ [111]<br>(0.88±0.4)x10$^{-18}$ [poly] | (1.1±0.4) x 10$^{-18}$ [poly] |
| GaSe | 2.1x10$^{-18}$ (10 µm) | [26] | (0.62±0.2)x10$^{-18}$ [100]<br>(0.61±0.2)x10$^{-18}$ [010] | (0.80±0.3) x 10$^{-18}$ [010] |
| ZGP | 5.5x10$^{-18}$ (2.2 µm) [010] | [27] | (1.9±0.6)x10$^{-18}$ [010]<br>(1.7±0.6)x10$^{-18}$ [101] | (1.7±0.7) x 10$^{-18}$ [010] |

**V. MPA COEFFICIENTS: THEORY VS. EXPERIMENT.**

.. The Keldysh model assumes the laser spectrum contains only one wavelength, i. e., has zero spectral width. The cited work ] contains estimations of corrections to the Keldysh model of nonlinear absorption related to non-zero spectral width.

We did quantum-mechanical calculation of the 6$^{th}$-order MPA coefficient ($\alpha_6$) in GaP using the Keldysh photoionization model [28]. GaP was chosen because its band-structure parameters are the best known among the four crystals considered here. In our four-band model of GaP, we considered contributions of heavy-hole, light-hole, and split-off valence bands, their nonparabolicity [29], and the local-field corrections. At the peak intensity of 100 GW/cm$^2$, the Keldysh parameter $\gamma = \sqrt{\frac{2m\omega^2 E_g}{e^2 E^2}}$ is between 2 and 3 for GaP, depending on the crystalline direction. Here $m$ is the reduced effective electron-hole mass, $\omega$ is the laser angular frequency, $E_g$ is the bandgap, $e$ is the electron charge, and $E \approx 5 \times 10^8$ V/m is the laser pulse electric field inside the sample. At these levels of the field, we neglected the coupling between the valence bands and estimate their contributions to the nonlinear absorption as independent. Considering the double degeneracy of the bands involved, using the effective masses from Ref. [29], and applying a modified form of the Keldysh formula reported in [30], we obtained the MPA coefficients for the three GaP orientations that are summarized in Table 3. Despite the broad laser spectrum, we found that the correction terms to the Keldysh model (that assumes a monochromatic radiation) are small and the Keldysh model is a good approximation of the MPA rate under our experimental conditions [31].

The experimental values of the MPA coefficients were obtained from the Z-scan data with 30 fs pulses, similar to Fig. 3(b), – by numerically solving Eq. (6) and neglecting FCA due to short pulse duration. We used two scenarios in our experimental data processing. In the first scenario, only the first two terms on the right side of Eq. (6) were taken into account, neglecting the third (self-focusing) term. The rationale for such an approximation was that (i) self-focusing plays a smaller role than that predicted by Eq. (6) since it may partly be offset by the negative refractive index contribution due to the free carriers, whose concentration $n_{FC}$ sharply increases as the beam self-focuses ($n_{FC} \sim w^{-12}$ for the 6$^{th}$-order MPA in GaP, where $w$ is the beam radius) and (ii) the nonlinear Schrödinger equation in the form of (6) neglects chromatic dispersion and is too approximate for the few-cycle pulse [32]. In the second scenario, we have included the 3$^{rd}$ term in Eq. (6) using our measured values for $n_2$ to incorporate self-focusing effects. The two sets of MPA coefficients corresponding to the two scenarios of experimental data processing are presented in Table. 3.The ZnSe results were not included since it is difficult to separate the contributions of the 5- and 6-photon MPAs, unless a laser with a different center wavelength is used. For GaSe and ZGP, the difference between extracted experimental $\alpha_4$ values for the two scenarios of data processing is within a factor of 3. However, for GaP, the extracted $\alpha_6$ values differ by a factor of ~ 50 for the two scenarios. This large difference is explained by a much higher sensitivity of the 6$^{th}$-order MPA to the beam size: for example, a 33% uncertainty of the beam size inside the sample leads to a 50-fold uncertainty in the extracted MPA coefficient. The theoretical $\alpha_6$ coefficients for GaP fit in between the experimental values of the 1$^{st}$ and 2$^{nd}$ scenario, being closer to the 1$^{st}$-scenario with no self-focusing included.

Table 3. MPA coefficients for GaP, GaSe, and ZGP: experiment ($\tau$=30-fs pulses) and theory.

| Crystal | Crystal orientation | MPA coefficient | Experiment (w/o self-focusing) | Experiment (with self-focusing) | Theory |
|---|---|---|---|---|---|
| GaP | [100] | $\alpha_6$ | 2.7x10$^{-10}$ cm$^9$/GW$^5$ | 5.2x10$^{-12}$ cm$^9$/GW$^5$ | 3.1x10$^{-10}$ cm$^9$/GW$^5$ |
|  | [110] |  | 1.1x10$^{-9}$ cm$^9$/GW$^5$ | 2.1x10$^{-11}$ cm$^9$/GW$^5$ | 3.4x10$^{-10}$ cm$^9$/GW$^5$ |
|  | [111] |  | 6.8x10$^{-10}$ cm$^9$/GW$^5$ | 1.3x10$^{-11}$ cm$^9$/GW$^5$ | 3.6x10$^{-10}$ cm$^9$/GW$^5$ |
| GaSe | [100] | $\alpha_4$ | 6.0x10$^{-7}$ cm$^5$/GW$^3$ | 2.1x10$^{-7}$ cm$^5$/GW$^3$ | - |
|  | [010] |  | 6.2x10$^{-7}$ cm$^5$/GW$^3$ | 2.2x10$^{-7}$ cm$^5$/GW$^3$ | - |
| ZGP | [010] | $\alpha_4$ | 1.1x10$^{-5}$ cm$^5$/GW$^3$ | 5.1x10$^{-6}$ cm$^5$/GW$^3$ | - |
|  | [101] |  | 5.2x10$^{-6}$ cm$^5$/GW$^3$ | 2.4x10$^{-6}$ cm$^5$/GW$^3$ | - |

## VI. Conclusion

Using femtosecond pulses at 2.35-μm we characterized multiphoton absorption, its anisotropy, and nonlinear refraction in GaP, ZnSe, GaSe, and ZGP crystals. For the first time to our knowledge MPA with $N>4$ was characterized in crystals. We found that the predominant MPA orders were $N = 6$ for GaP, $N =5$ for ZnSe, and $N =4$ for GaSe and ZGP, in agreement with their bandgap energies and the perturbation model. By varying the pulse length while maintaining a constant peak intensity, we observed a strong dependence of the nonlinear absorption on pulse duration. We found that the contribution of accumulated free carriers is much higher than what would be expected from their linear absorption cross section, suggesting that high-field effects such as intraband scattering and absorption to the higher lying bands may play an important role. While Table 2 provides a guidance in terms of scaling laws and intensities corresponding to the onset of MPA, a better model is needed to extract MPA coefficients from experimental data. The model should adequately describe self-focusing by adding a dispersion term to the nonlinear Schrödinger equation and the contribution of generated free carriers to the refractive index and absorption, taking into account strong-field effects; however, this is beyond the scope of this article. In view of the growing use of ultrafast high peak intensity MIR lasers, our results provide a basis to evaluate the impact of high-order effects in diverse applications.

1. S. Ghimire and D. A. Reis, High-harmonic generation from solids, Nature Phys. 15, 10 (2019).

2. O. Schubert, M. Hohenleutner, F. Langer, B. Urbanek, C. Lange, U. Huttner, D. Golde, T. Meier, M. Kira, S. W. Koch and R. Huber, Sub-cycle control of terahertz high-harmonic generation by dynamical Bloch oscillations, Nature Photon. 8, 119 (2014).

3. A. A. Lanin, E. A. Stepanov, A. B. Fedotov, and A. M. Zheltikov, Mapping the electron band structure by intraband high-harmonic generation in solids, Optica 4, 516 (2017).

4. M. R. Shcherbakov, H. Zhang, M. Tripepi, G. Sartorello, N. Talisa, A. AlShafey, Z. Fan, J. Twardowski, L. A. Krivitsky, A. I. Kuznetsov, E. Chowdhury, and G. Shvets, Generation of even and odd high harmonics in resonant metasurfaces using single and multiple ultra-intense laser pulses, Nature Commun. 12, 4185 (2021).

5. Y. Yu, X. Gai, P. Ma, D.Y. Choi, Z.Y. Yang, R.P. Wang, S. Debbarma, S.J. Madden, B. Luther-Davies, A broadband, quasi-continuous, mid-infrared supercontinuum generated in a chalcogenide glass waveguide, Laser Photon. Rev. 8, 792 (2014).

6. O. Mouawad, P. Béjot, F. Billard, P. Mathey, B. Kibler, F. Désévédavy, G. Gadret, J.-C. Jules, O. Faucher, F. Smektala, Filament-induced visible-to-mid-IR supercontinuum in a ZnSe crystal: Towards multioctave supercontinuum absorption spectroscopy, Opt. Mater. 60, 355 (2016).

7. R. Suminas, G. Tamošauskas, G. Valiulis, V. Jukna, A. Couairon, and A. Dubietis, Multi-octave spanning nonlinear interactions induced by femtosecond filamentation in polycrystalline ZnSe, Appl. Phys. Lett. 110, 241106 (2017).

8. S. Vasilyev, I. Moskalev, V. Smolski, J. Peppers, M. Mirov, A. Muraviev, K. Vodopyanov, S. Mirov, and V. Gapontsev, Multi-octave visible to long-wave IR femtosecond continuum generated in Cr:ZnS-GaSe tandem, Opt. Express 27, 16405 (2019).

9. N. Leindecker, A. Marandi, R.L. Byer, K. L. Vodopyanov, J. Jiang, I. Hartl, M. Fermann, and P. G. Schunemann, Octave-spanning ultrafast OPO with 2.6-6.1 μm instantaneous bandwidth pumped by femtosecond Tm-fiber laser, Opt. Express 20, 7047-7053 (2012).

10. Q. Ru, T. Kawamori, P. G. Schunemann, S. Vasilyev, S. B. Mirov, and K. L. Vodopyanov, Two-octave-wide (3–12 μm) subharmonic produced in a minimally dispersive optical parametric oscillator cavity, Optics Letters 46, 709 (2021).

11. S. Vasilyev, I. S. Moskalev, V. O. Smolski, J. M. Peppers, M. Mirov, A. V. Muraviev, K. Zawilski, P. G. Schunemann, S. B. Mirov, K. L. Vodopyanov, and V. P. Gapontsev, Super-octave longwave mid-infrared coherent transients produced by optical rectification of few-cycle 2.5- μm pulses, Optica 6, 111 (2019).

12. J. Zhang, K. Fritsch, Q. Wang, F. Krausz, K. F. Mak, and O. Pronin, Intra-pulse difference-frequency generation of mid-infrared (2.7-20 μm) by random quasi-phase-matching, Opt. Lett. 44, 2986 (2019).

13. B. Monoszlai, P. S. Nugraha, Gy. Tóth, Gy. Polónyi, L. Pálfalvi, L. Nasi, Z. Ollmann, E. J. Rohwer, G. Gäumann, J. Hebling, T. Feurer, and J. A. Fülöp, Measurement of four-photon absorption in GaP and ZnTe semiconductors, Opt. Express 28, 12352 (2020).

14. M. Sheik-Bahae, A. A. Said, T. H. Wei, D. J. Hagan, and E. W. van Stryland, Sensitive measurement of optical nonlinearities using a single beam, IEEE Journal of Quantum Electronics **26**, 760 (1990).

15. D.K. Schroder, "*Semiconductor Material and Device Characterization, 3rd ed.*" (Wiley-IEEE, 2015).

16. A. Couairon, E. Brambilla, T. Corti, D. Majus, O. de, and M. Kolesik, Practitioner's guide to laser pulse propagation models and simulation, European Physical Journal: Special Topics 199, 5 (2011).

17. P. G. Schunemann, K. T. Zawilski, L. A. Pomeranz, D. J. Creeden, and P. A. Budni, Advances in nonlinear optical crystals for mid-infrared coherent sources, J. Opt. Soc. Amer. 33, D36 (2016).

18. Q. Ru, N. Lee, X. Chen, K. Zhong, G. Tsoy, M. Mirov, S. Vasilyev, S. B. Mirov, and K. L. Vodopyanov, Optical parametric oscillation in a random polycrystalline medium, Optica 4, 617-618 (2017).

19. S. Vasilyev, I. Moskalev, V. Smolski, J. Peppers, M. Mirov, A. Muraviev, K. Vodopyanov, S. Mirov, and V. Gapontsev, Multi-octave visible to long-wave IR femtosecond continuum generated in Cr:ZnS-GaSe tandem, Opt. Express 27, 16405 (2019).

20. W. G. Spitzer, M. Gershenzon, C. J. Frosch, and D. F. Gibbs, Optical absorption in n-type gallium phosphide, J. of Physics and Chemistry of Solids, 11, 339-341 (1959).

21. B. V. Dutt, O. K. Kim, and W. G. Spitzer, Free-carrier absorption of n-type ZnSe:Al, J. Appl. Phys. 48, 2110 (1977).

22. A. Martin, S. Combrié, A. de Rossi, G. Beaudoin, I. Sagnes, and F. Raineri, Nonlinear gallium phosphide nanoscale photonics Invited., Photon. Res. 6, B43-B49 (2018).

23. T. R. Ensley and N. K. Bambha, Ultrafast nonlinear refraction measurements of infrared transmitting materials in the mid-wave infrared, Optics Express 27, 37940 (2019).

24. G. N. Patwardhan, J. S. Ginsberg, C. Y. Chen, M. M. Jadidi, and A. L. Gaeta, Nonlinear refractive index of solids in mid-infrared, Optics Letters 46, 1824 (2021).

25. K. Werner, M. G. Hastings, A. Schweinsberg, B. L. Wilmer, D. Austin, C. M. Wolfe, M. Kolesik, T. R. Ensley, L. Vanderhoef, A. Valenzuela, and E. Chowdhury, Ultrafast mid-infrared high harmonic and supercontinuum generation with $n_2$ characterization in zinc selenide, Opt. Express 27, 2867 (2019).

26. J. J. Pigeon, D. A. Matteo, S. Ya. Tochitsky, I. Ben-Zvi, and C. Joshi, Measurements of the nonlinear refractive index of $AgGaSe_2$, GaSe, and ZnSe at 10 μm, J. Opt. Soc. Am. B 37, 2076 (2020).

27. W. Li, Y. Li, Y. Xu, J. Lu, P. Wang, J. Du, and Y. Leng, Measurements of nonlinear refraction in the mid-infrared materials $ZnGeP_2$ and $AgGaS_2$, Applied Physics B: Lasers and Optics **123**, 82 (2017).

28. L.V. Keldysh, Ionization in the field of a strong electromagnetic wave, Sov. Phys. – JETP 20, 1307-1314 (1965).

29. I. Vurgaftman, J. R. Meyer, L. R. Ram-Mohan, Band parameters for III-V compound semiconductors and their alloys, J. Appl. Phys. 89 (11), 5815-5875 (2001).

30. V. Gruzdev, Fundamental mechanisms of laser damage of dielectric crystals by ultrashort pulse: ionization dynamics for the Keldysh model, Opt. Eng. 53(12), 122515 (2014).

31. V. Gruzdev, O. Sergaeva, Simulation of the photoionization of non-metal crystals by few-cycle femtosecond laser pulses, Opt. Eng. 61(2), 021006 (2022).

32. Miroslav Kolesik, College of Optical Sciences, Univ. Arizona, Tucson, USA (private communication).